\documentclass[]{spie}  

\usepackage{amsmath,amsfonts,amssymb}
\usepackage{graphicx}
\usepackage[colorlinks=true, allcolors=blue]{hyperref}

\title{Simulations of athermal phonon propagation in a cryogenic semiconducting bolometer}

\author[a,b]{S. L. Stever}
\author[c]{F. Couchot}
\author[d]{B. Maffei}
\affil[a]{Okayama University, 3-1-1 Tsushimanaka, Kita-ku, Okayama 700-0082 Japan}
\affil[b]{Kavli Institute for the Physics and Mathematics of the Universe (WPI),The University of Tokyo Institutes for Advanced Study, The University of Tokyo, Kashiwa, Chiba 277-8583 Japan}
\affil[c]{Laboratoire de Physique des 2 Infinis Irène Joliot Curie, 15 Rue Georges Clemenceau, 91400 Orsay, France}
\affil[c]{Institut d'Astrophysique Spatiale, Bâtiment 121, 91405 Orsay France}

\authorinfo{Further author information: (Send correspondence to S. L. Stever)\\S. L. Stever: E-mail: sstever@okayama-u.ac.jp}

\begin{document} 
\maketitle

\begin{abstract}
We present three Monte Carlo models for the propagation of athermal phonons in the diamond absorber of a composite semiconducting bolometer `Bolo 184'. Previous measurements of the response of this bolometer to impacts by $\alpha$ particles show a strong dependence on the location of particle incidence, and the shape of the response function is determined by the propagation and thermalisation of athermal phonons. The specific mechanisms of athermal phonon propagation at this time were undetermined, and hence we have developed three models for probing this behaviour by attempting to reproduce the statistical features seen in the experimental data. The first two models assume a phonon thermalisation length determined by a mean free path $\lambda$, where the first model assumes that phonons thermalise at the borders of the disc (with a small $\lambda$) and the second assumes that they reflect (with a $\lambda$ larger than the size of the disc). The third model allows athermal photons to propagate along their geometrical line of sight (similar to ray optics), gradually losing energy. We find that both the reflective model and the geometrical model reproduce the features seen in experimental data, whilst the model assuming phonon thermalisation at the disc border produces unrealistic results. There is no significant dependence on directionality of energy absorption in the geometrical model, and in the schema of this thin crystalline diamond, a reflective absorber law and a geometrical law both produce consistent results.
\end{abstract}


\section{INTRODUCTION}
\label{sec:intro}  

Characterisation and simulation of the response functions of cryogenic bolometers is often dependent upon the geometry and thermal architecture of the specific detector. Typical cryogenically-cooled semiconducting bolometers employ a thermistor composed of a heavily-doped semiconducting material in thermal contact with a much larger crystalline absorber. Characterising the response function of such detectors depends not only on understanding its steady-state response (i.e. IV characteristics) but also on the specific mechanisms of energy deposition and propagation between the detector materials.\\

In this manuscript, we present a follow-up study on prior measurements of a semiconducting bolometer `Bolo 184' in response to energy deposition by $\alpha$ particles at 100 - 200 mK. Our previous measurements \cite{stever2019characterisation}$^{,}$\cite{stever2019towards} as well as impulse response modelling \cite{stever2018new} have shown a strong position-dependent effect in relation to the location of the $\alpha$ particle impact on the detector absorber. Upon absorption of the $\alpha$ particle, its energy is translated into athermal phonons which thermalise into thermal phonons, after which the energy propagation is diffusive. \\

Whilst the majority of the thermal properties of this detector have been described or devised in prior work, the specific attributes of the athermal phonon propagation, such as its mean free path $\lambda$, are not known. Thus, our objective is to exploit simple physical modelling to attempt to reproduce the macroscopic statistical attributes of our experimental data in order to form a better understanding of these mechanisms.\\

In this manuscript, we will briefly describe the bolometer and experimental outcomes motivating the modelling in Section \ref{sec:bolo}, describe the contruction of three athermal phonon propagation models in Section \ref{sec:sims}, and the results in Section \ref{sec:results}.\\

\section{Bolo 184 and experimental outcomes}
\label{sec:bolo}

Bolo 184 is a composite semiconductor bolometer originally developed for the DIABOLO photometer mounted on the Infrared Testa Grigia Observatory for observing diffuse millimetre-wave astronomical sources \cite{benoit2000calibration}\!. The series of experiments of interest for this manuscript were conducted between 2015 and 2019 on one spare bolometer, irradiated by a $^{241}$Am $\alpha$ particle source at 200 mK. The experiments have been described in detail in prior work \cite{stever2019characterisation}$^{,}$\cite{stever2019towards} and so they will be described only briefly here.\\

   \begin{figure} [ht]
   \begin{center}
   \begin{tabular}{c} 
   \includegraphics[width=0.3\textwidth]{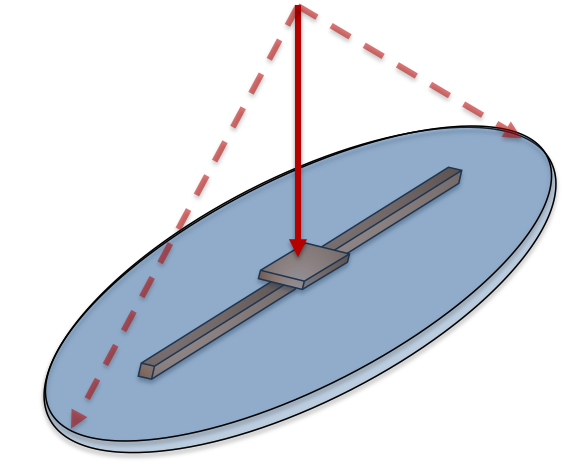}
   \end{tabular}
   \end{center}
   \caption{Simplified diagram of Bolo 184. \textit{Blue}: Diamond absorber (semi transparent). \textit{Gray}: NTD Germanium thermistor. \textit{Red}:  Travel direction of incident alpha particles. The coupling to the 200 mK thermal bath is at both ends of the thermistor, but is not shown in this diagram.}
   { \label{fig:b184diagram} }
 \end{figure} 
   
Bolo 184 is composed of several layers. The thermistor is composed of Neutron Transmutation Doped (NTD) germanium, with two long and thin legs coupled to the thermal bath. The centre of the NTD thermistor is glued below an absorber constructed of diamond (type 2A). It is important to note that the absorber itself is not coupled to the thermal bath directly, and hence all energy deposited into it must eventually flow through the thermistor where it is read as signal. A simplified diagram of Bolo 184 is shown in Figure \ref{fig:b184diagram}. We have omitted a sapphire layer below the thermistor, which acts as mechanical support, as well as a thin layer of deposited bismuth on the top side of the absorber disc.\\

\begin{figure} [ht]
   \begin{center}
   \begin{tabular}{c} 
   \includegraphics[width=0.8\textwidth]{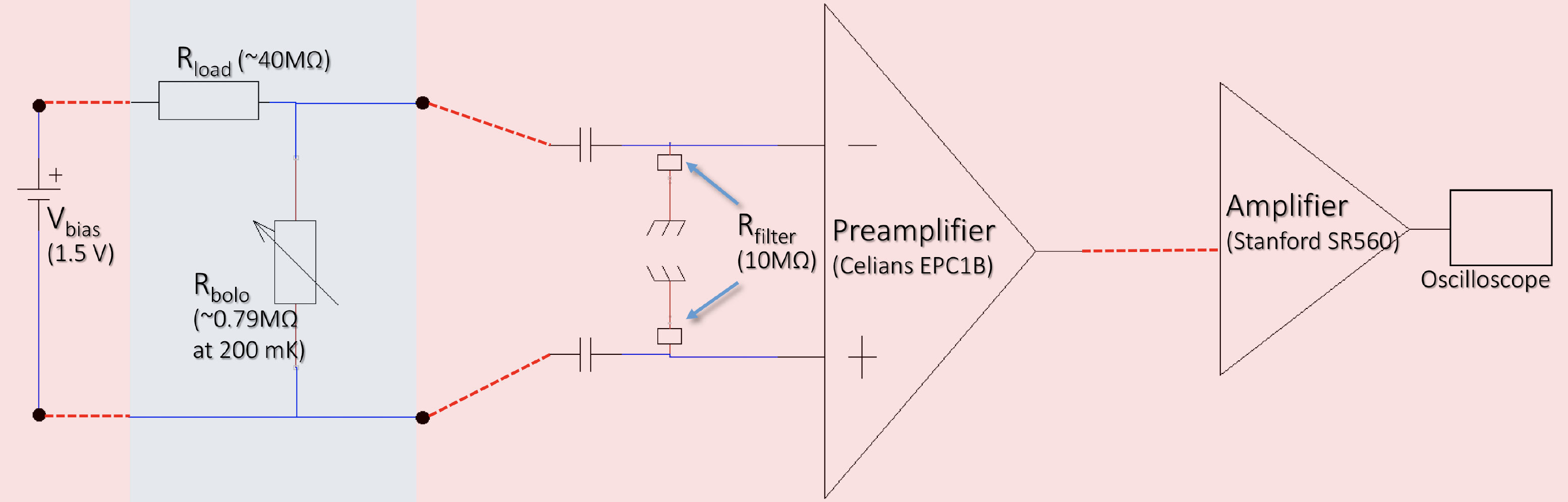}
   \end{tabular}
   \end{center}
   \caption{Diagram of the bolometer circuit for the experiments described in this manuscript, showing cold (in-cryostat, blue) and warm (external, red) components.}
   { \label{fig:b184circuit} }
 \end{figure}

As in previous manuscripts \cite{stever2019characterisation}$^{,}$\cite{stever2019towards}\!, we situate a $^{241}$Am $\alpha$ particle source directly above the bolometer, inside of a conical horn which acts as a diaphragm to limit the $\alpha$ propagation to within the geometrical area of the detector. The bolometer is positioned at an oblique angle relative to the horn. The circuit contains a
$\approx$40 ${\rm M}\Omega$ resistor, and the bolometer is voltage-biased at 1.5 V. The cold experimental apparatus is cooled to 200 mK using an inverted dilution refrigerator and held at 200 $\pm$ 0.1 mK by a resistor heater and PID. The complete circuit is shown in Figure \ref{fig:b184circuit}. Further details about the bolometer or the experimental setup of this manuscript are available in previous publications \cite{stever2019towards}\,.\\

\begin{figure} [ht]
   \begin{center}
   \begin{tabular}{c} 
   \includegraphics[width=0.5\textwidth]{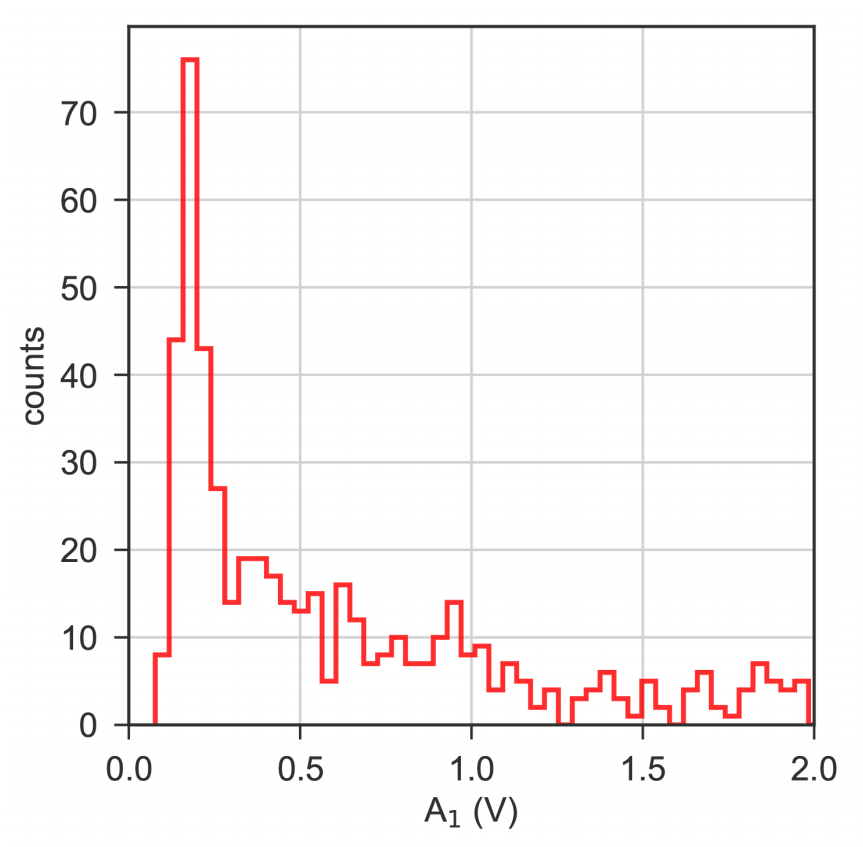}
   \end{tabular}
   \end{center}
   \caption{Histogram of the fast amplitude component $A_{1}$ from 200 mK Bolo 184 data. Taken from\cite{stever2019characterisation}\!.}
   { \label{fig:A1data} }
 \end{figure}

Our aim is to reproduce a statistical distribution of the results from prior publication\cite{stever2019towards}\!.
Specifically, we aim to reproduce the distribution of the 1$^{\rm st}$ amplitude component, which is the fast component arising from the propagation of athermal phonons. This distribution is shown in Figure \ref{fig:A1data}. Most of the incident $\alpha$ particles will impact the periphery of the absorber, creating low 1$^{\textrm{st}}$ amplitudes owing to a lower portion of the total deposited energy arriving quickly at the central absorber; in this case, the main driver for the pulse shape is thermal diffusion from normal phonon propagation. This phenomenon explains the strong skew towards low amplitudes in the distribution. \\

In order to attempt to reproduce this distribution, we present three Monte Carlo models for the propagation of energy in the diamond absorber of Bolo 184. In each model we choose an impact location on the absorber disc and generate a large number of particles which represent athermal phonons. \\
 
In our prior work, we assumed that athermal phonons thermalise at the disc borders or in the nearest Bi interface they encounter in the absorber, and that their thermalisation distance is described by an exponential probability relating to their mean free path. We will now examine (\textit{1}) whether this assumption a can reproduce the defining characteristics of these pulses, (\textit{2}) whether a reflective phonon schema, in which the phonons reflect off disc borders with a mean free path larger than the disc radius, or (\textit{3}) whether the distance of athermal phonon thermalisation follows a simple geometrical law, i.e. a geometrical line-of-sight to the central region after reflecting on the disc borders (analogous to ray tracing in optics). We will also use this to determine the mean free path $\lambda$ of athermal phonons in the absorber. The `edge absorption model', `reflective absorber model', and the `geometrical absorber model' will be described in the following section.\\

\section{Description of absorber models}
\label{sec:sims}

We will first describe the edge absorption model, followed by the minor modifications to produce the reflective absorber model. Finally, the geometrical absorber schema will be described.\\

\subsection{Edge absorption model}
The edge absorption model assumes athermal phonon propagation and thermalisation in a single timestep. The athermal phonon thermalisation function loops over $N$ athermal phonons per alpha particle where $N\approx$100000. We take a disc of equal size to the diamond-bismuth absorber ($r_{\textrm{disc}}$ = 1.75 mm). For each $\alpha$ particle, we generate a random $\rho$ with a linear probability and we define a $\phi$ with a flat probability between 0 and 2$\pi$, corresponding to a uniform distribution of the $\alpha$'s impact on the disk. We determine the $\alpha$ particle starting location using that $\rho$ and $\phi$.\\

In the main loop, the position of the $\alpha$ particle in $x$ and $y$ is simply its impact point. All athermal phonons belonging to that $\alpha$ particle (10000 phonons per $\alpha$) start from that initial position. For each phonon, we choose a random polar angle $\theta_{\textrm{phonon}}$ with respect to the disc perpendicular $z$, and a random $\phi_{\textrm{phonon}}$ between 0 and 2$\pi$, giving a uniform spherical distribution to the phonon 3D direction, and a projected path length $\rho_{\textrm{phonon}}$ which follows an exponential law:

\begin{equation}
\label{rho_phonon}
\rho_{\textrm{phonon}} = -\lambda \cdot \sin(\theta_{\textrm{phonon}}) \cdot \log(\texttt{random})
\end{equation}

The phonon is propagated by the distance $\rho_{\textrm{phonon}}$ in the direction $\phi_{\textrm{phonon}}$. For simplicity, we simulate the thermistor as a circle of $r_{\textrm{sens}}$ = 600 $\mu$m - the actual sensor is 800 $\mu$m $\times$ 260 $\mu$m in size, so we would assume the size of the disc in good thermal contact with the sensor to be roughly this size (or larger, depending on the strength of the thermal coupling). 
If the phonon's new distance is less than $r_{\textrm{sens}}^{2}$, the phonon counts toward the count of the total number of phonons arriving in the area with good thermal contact with the sensor, i.e. having a strong athermal (fast) pulse.\\

In the edge absorption model, we assume that $\lambda$ is small compared with the disc radius (with $\lambda = 20$\% $r_{\textrm{disc}}$). \\

\begin{figure}
  \centering
  \includegraphics[width=0.4\textwidth]{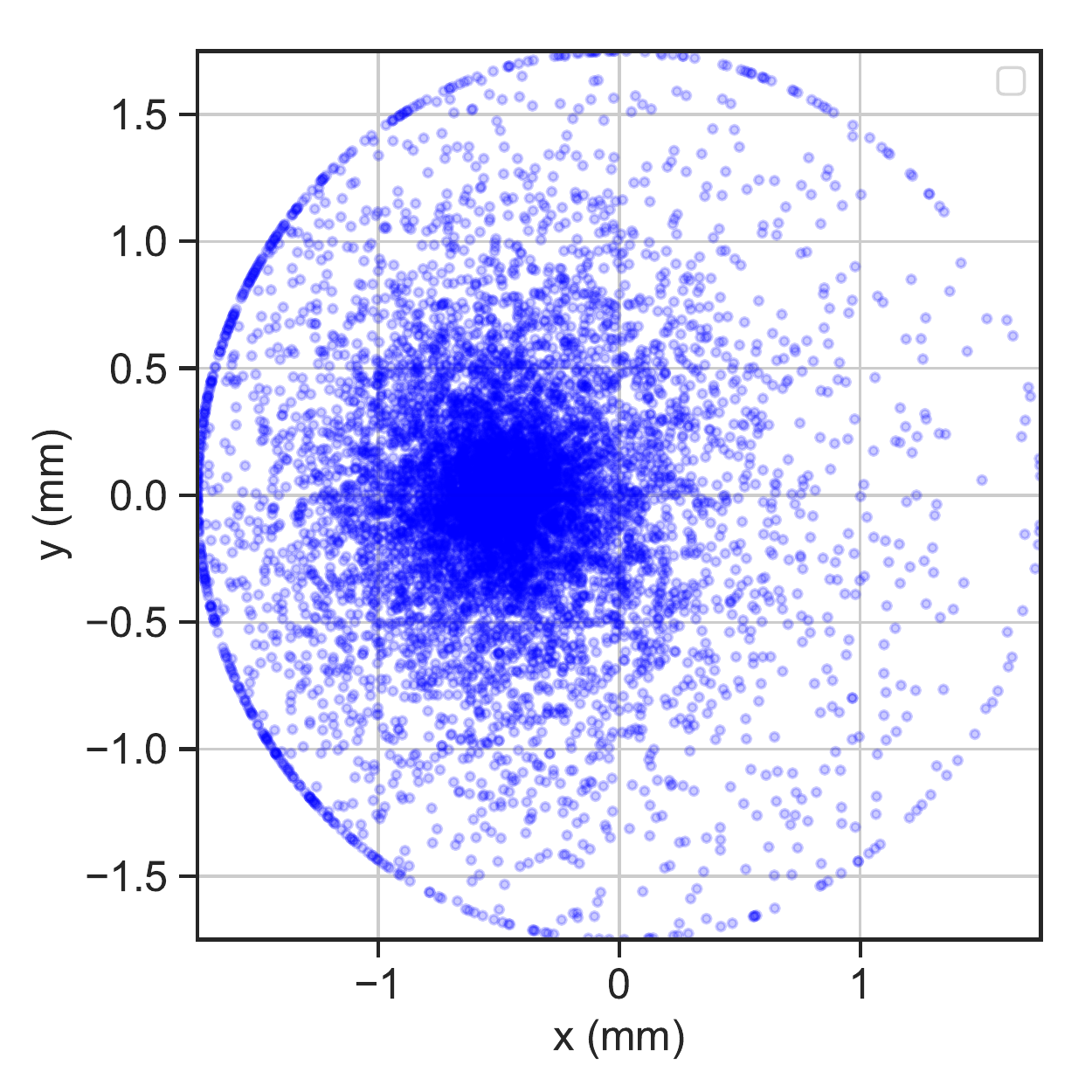}
      \caption{Output of the edge absorption model for one $\alpha$ particle, showing athermal phonon thermalisation locations with an initial particle impact point in the negative $x$ region of the disc. Single phonon thermalisation points are shown as blue dots.}
      \label{img:BPmodel}
\end{figure}

The loop calculates whether the total displacement of each particle falls outside the border of the disc edge, and places the particle location at the border if it does. Otherwise, the particle thermalises in the location calculated by $\phi$ and $\rho_{\textrm{phonon}}$. Typical single-$\alpha$ results are shown in Figure~\ref{img:BPmodel}, where we show the phonon thermalisation pattern for an $\alpha$ particle impact point at the negative $x$ region of the absorber disc. This image demonstrates the strong thermalisation at the disc edges, behaviour which would scale with increased $\lambda$.\\

For the purposes of analysis, for all models described in this manuscript, a large number of $\alpha$ particles (evenly distributed across the disc) propagate as a larger amount of athermal phonons, in order to determine macroscopic statistical properties.\\

\subsection{Reflective absorber model}
This model is similar to the edge absorption model, with the exception that $\lambda$ is now assumed to be much larger than the disc size ($\approx$3$\times r_{\textrm{disc}}$) and that phonons impacting the disc border have reflective behaviour. \\

If the phonon falls outside $r_{\textrm{disc}}$, it reflects off the border. We find the square of its movement distance,\newline $d^{2}$ = $r_{\textrm{disc}}^{2} - \rho^{2}$, and translate the coordinates of its movement using $b' = x \cdot \cos(\phi) + y \cdot \sin(\phi)$. We find the new distance travelled by the phonon, $s = -b' + \sqrt{b'^{2} + d^{2}}$. If it is less than $\rho_{\textrm{phonon}}$, we calculate the $x$ and $y$ locations where the phonon intersects with the border:

\begin{equation}
\label{balistick2}
x_{\textrm{border}} = x_{\alpha} + s \cdot \cos(\phi_{\textrm{phonon}}) \\ y_{\textrm{border}} = y_{\alpha} + s \cdot \sin(\phi_{\textrm{phonon}})
\end{equation}

We define the $\phi_{\textrm{border}}$ of the newly-reflected phonon by taking the arctangent of the $x$ and $y$ border interaction locations, and invert the angle using $\phi_{\textrm{reflect}} = 2 \cdot \phi_{\textrm{border}} - \phi_{\textrm{phonon}} + \pi$. We subtract $s$ from $\rho_{\textrm{phonon}}$ and use it to find the translated phonon location:

\begin{gather}
\label{balistick3}
x_{\textrm{trans}} = x_{\textrm{border}} + \rho_{\textrm{phonon}} - s \cdot \cos(\phi_{\textrm{reflect}}) \\ y_{\textrm{trans}} = y_{\textrm{border}} + \rho_{\textrm{\textrm{phonon}}} - s \cdot \sin(\phi_{\textrm{reflect}})
\end{gather}

The new distance is calculated in the usual way using $d = \sqrt{x_{\textrm{trans}}^{2} + y_{\textrm{trans}}^{2}}$. We check again whether $d < r_{\textrm{sens}}$, and add it to the total number of athermal phonons caught in the central radius if this is the case. If not, the process is repeated if the new distance of the phonon falls outside the edges of the absorber again. The phonon continues to reflect off the borders in this way until it is either caught within $r_{\textrm{disc}}$ or it thermalises in an area which is $<$ $r_{\textrm{disc}}$ but $> r_{\textrm{sens}}$, when its total path length would exceed
$\rho_{\textrm{phonon}}$.\\

\subsection{Geometrical absorber model}
The geometrical absorber model is intended to simulate geometrical effects with a progressive phonon energy loss (instead of a total energy loss after elastic reflections, and an exponential
random path distribution). It is similar to the reflective absorber model, except that $\rho$ is not a random variable. Similarly to the last case, the $x$ and $y$ locations of each phonon are continuously updated, with an additional component in $z$, and when their $d = \sqrt{x^{2}+y^{2}}$ is $>r_{\textrm{disc}}$, they are reflected back into the disc. At each reflection on the border or the bottom of the disc (but not the top of the diamond, which is polished and assumed to be `mirror-like'), the phonon is assumed to lose a proportion of energy into the disc. Each phonon continues to propagate until it loses all of its energy, or until it makes 1000 reflections. The code then records the total amount of energy deposited inside the central radius relative to outside of it. \\

There are a few options for how much energy a phonon loses into the disc with each impact: the energy loss per reflection, $\epsilon$, can be a flat amount $E_{\textrm{dep}}$ = $\epsilon$; an energy deposition which is maximum when parallel to the Bi layer (bottom of the disc) where $E_{\textrm{dep}}$ = $\epsilon \cdot (1 + \cos^{2}(\theta)$) / 2; energy deposition which favours perpendicularity with the Bi layer $E_{\textrm{dep}}$ = $\epsilon$ $\cdot (\cos^{2}(\theta))$. The mechanisms of phonon energy loss into the diamond crystal may be anisotropic due to crystal lattice structure, the exact mechanics of which are not presently available; to this end, we will attempt several possibilities, and we will examine each of these, denoting them `flat', `absorption', and `perpendicularity' (respectively).\\

\section{Modelling results}
\label{sec:results}
In this section we will outline the results of the edge absorption, reflective absorber, and geometrical models, comparing their phonon thermalisation patterns and their ability to reproduce the distribution of the fast amplitude $A_{1}$.\\

\subsection{Edge absorption model - results}

   \begin{figure} [ht]
   \begin{center}
   \begin{tabular}{c} 
   \includegraphics[width=0.4\textwidth]{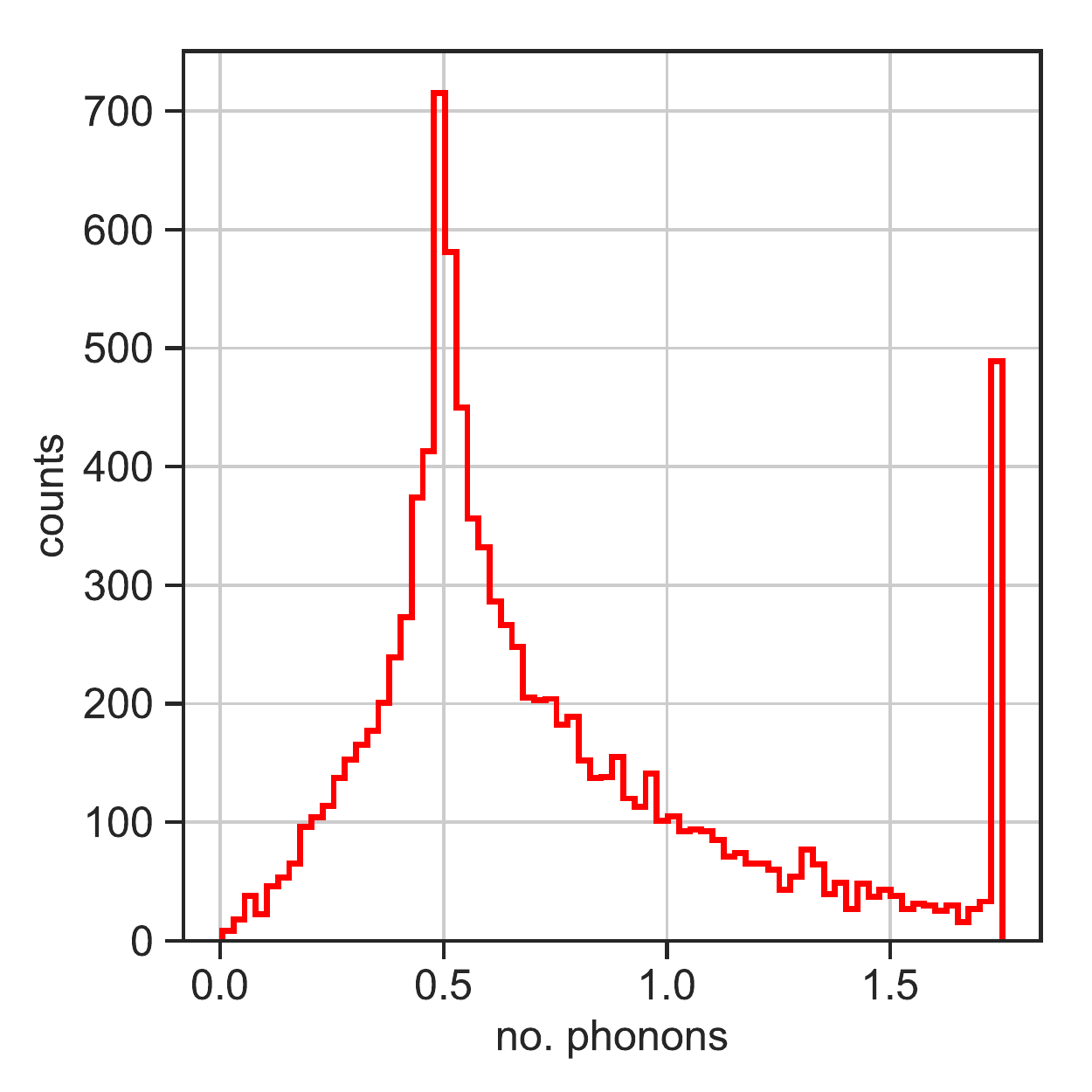}
   \end{tabular}
   \end{center}
   \caption{Histogram of the simulated `fast' amplitude $A_{1}$ produced by the edge absorption model.}
   { \label{fig:A1edge} }
 \end{figure} 
 
We compare the proportion of athermal phonons thermalised in the central sensor area, analogous to a the distribution of the `fast' amplitude in the pulses $A_{1}$. What we are attempting to reproduce is simply the shape of the statistical distribution shown in Figure \ref{fig:A1data}. For this model, the simulated amplitudes shown in Figure \ref{fig:A1edge} demonstrate a strong deviation from the data attributes, with a majority of the thermalised phonons appearing at the edge of the sensor as well as a significant skewness at the first peak.\\

\subsection{Reflective absorber model - results}

   \begin{figure} [ht]
   \begin{center}
   \begin{tabular}{c} 
   \includegraphics[width=0.5\textwidth]{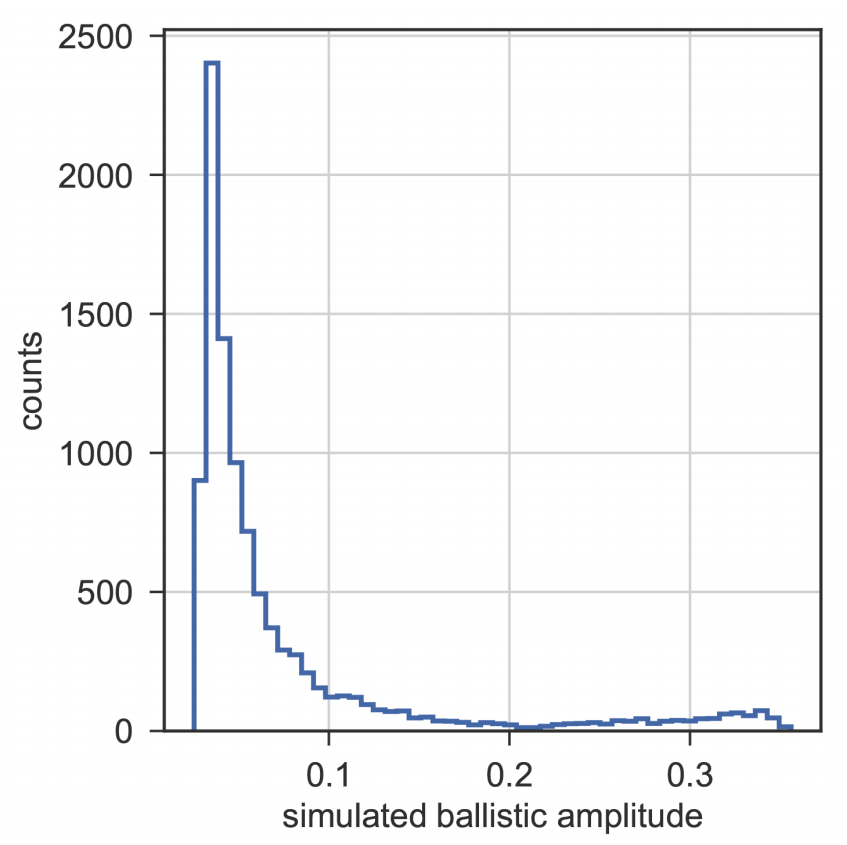}
   \end{tabular}
   \end{center}
   \caption{Histogram of the simulated `fast' amplitude $A_{1}$ produced by the reflective absorber model.}
   { \label{fig:A1refsim} }
 \end{figure} 
 
Repeating the same treatment as the previous section, we find that he shape of the distribution of simulated $A_{1}$ for the reflective absorber model qualitatively reproduces that of the first amplitude, with a similar first peak and with a small buildup of events at high `amplitudes' coming from far-away athermal phonon reflection on the borders. This simulated distribution is shown in Figure \ref{fig:A1refsim}.\\

\begin{figure}
  \centering
  \caption{Four random athermal phonon thermalisation patterns for $N_{\textrm{phonon}}$ = 10000, for one $\alpha$ particle in the reflective absorber model. Blue dots are the points in the disc where the ballistic phonon has deposited its energy.}
  \includegraphics[width=0.95\textwidth]%
    {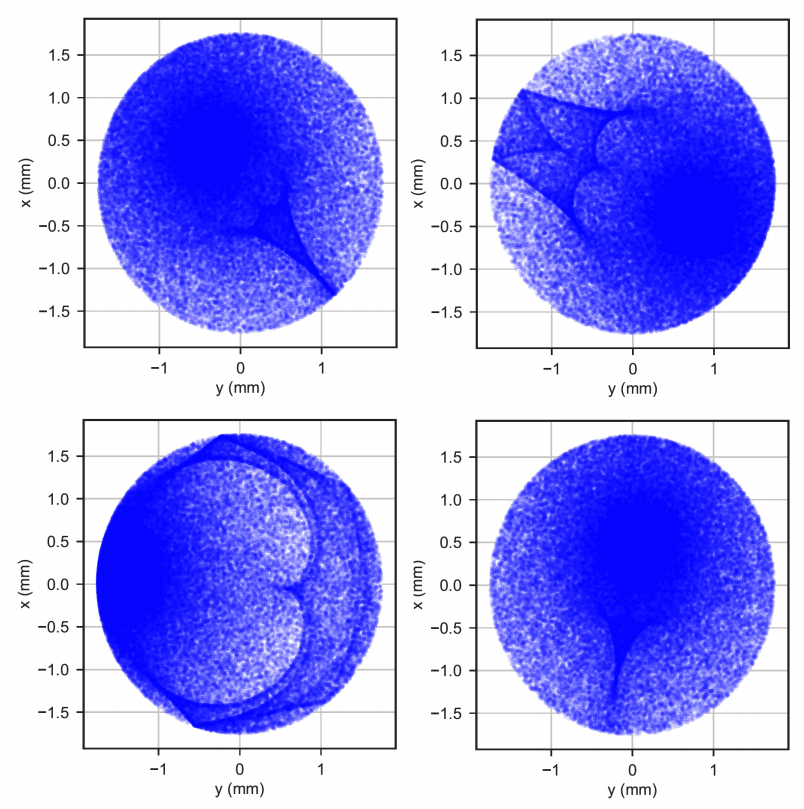}
  \label{fig:reflectivepattern}
\end{figure}

By plotting the $x$ and $y$ locations of phonon thermalisation within the disc surface in Figure \ref{fig:reflectivepattern}, we see clear areas where the phonons deposit their energy the most often, which change based on where the $\alpha$ particle originated. These images are not overwhelmingly useful unto themselves, but they do show that athermal phonons, under the assumption of a probabilistic mean free path and reflective properties at disc borders, tend to have geometrical distributions which clearly reflect the curvature of the disc edges in where they appear to thermalise.\\

Finally, we find the most consistent results when we assume a mean free path $\lambda$ of $\approx$ 3$\times$the geometrical size of the diamond absorber. This is in contrast with the edge absorption model, where a larger $\lambda$ only results in most athermal phonons thermalising in the disc border and removing the effective position-dependent effects. \\

\subsection{Geometrical absorber model - results}
\begin{figure}
  \centering
  \caption{Simulated athermal phonon amplitudes from the geometrical approximation using $r_{\textrm{sens}}$ = 600 $\mu$m with \textit{Left/blue}: absorption law $E_{dep}$;  \textit{Centre/green}: $E_{\textrm{dep}}$ with perpendicularity law; \textit{Right/pink}: flat $E_{\textrm{dep}}$.}
  \includegraphics[width=0.99\textwidth]%
    {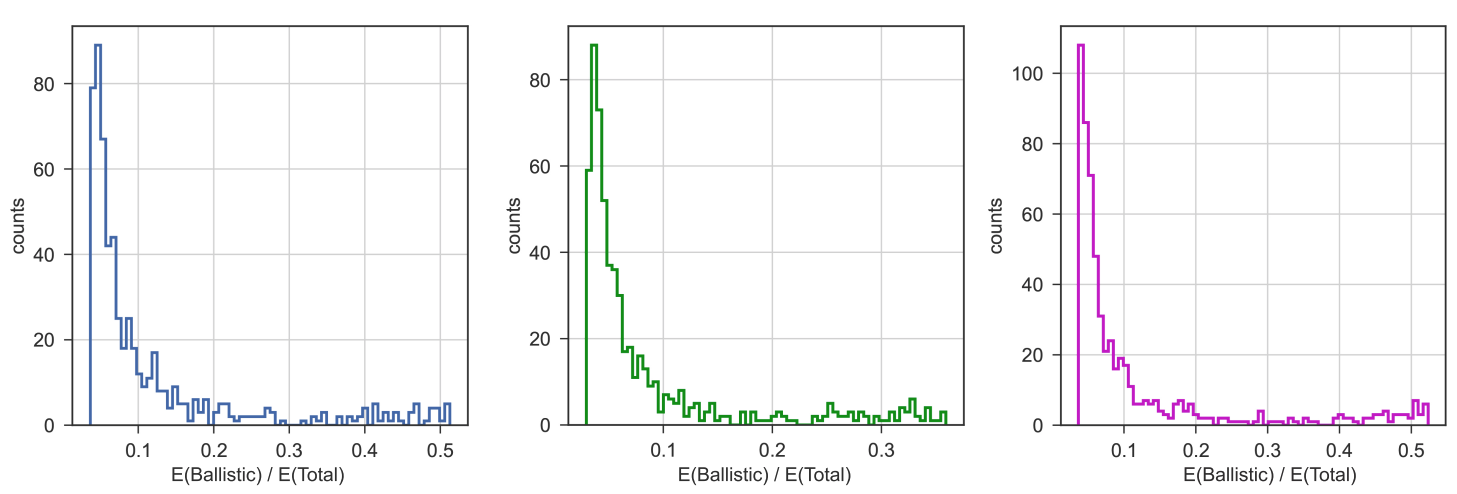}
  \label{img:titi_ballistic}
\end{figure}

Comparing the simulated amplitudes with those from the data, as we have done previously, we find in Figure~\ref{img:titi_ballistic} that the shape of the distribution is again similar for a $r_{\textrm{sens}}$ of 600 $\mu$m, consistent with the size of the actual thermistor. The numerical value (non-normalised in this case) represents the ratio of energy distributed inside vs. outside $r_{\textrm{sens}}$. The absorptive and perpendicular distributions appear to be very similar, and produce output which more closely resembles the actual amplitude distribution of the data; the flat energy deposition is slightly too peaked. \\

\begin{figure}
  \centering
  \caption{Three random athermal phonon thermalisation patterns for $N_{\textrm{phonon}}$ = 1, for one $\alpha$ particle. Dots are locations of phonons which have deposited some of their energy, lines are their paths as they reflect, and red stars are the $\alpha$ particle impact position. \textit{Top/blue}: $E_{\textrm{dep}}$ = $\epsilon \cdot (1 + \cos^{2}(\theta)$) / 2;  \textit{Centre/green}: $E_{\textrm{dep}}$ = $\epsilon$ $\cdot (\cos^{2}(\theta))$;  \textit{Bottom/pink}: flat $E_{\textrm{dep}}$.}
  \includegraphics[width=0.99\textwidth]%
    {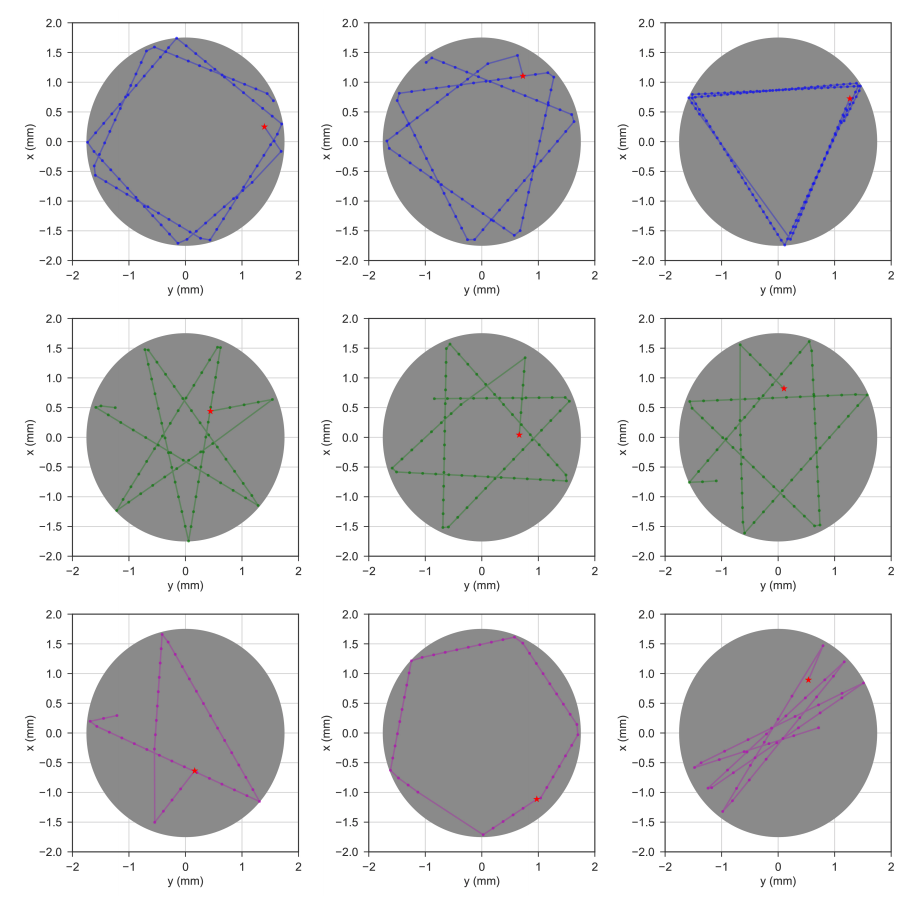}
  \label{img:titi_phononpattern}
\end{figure}

The phonon thermalisation patterns for this model are very different to the last case, with some of the phonon paths containing many reflections (Figure~\ref{img:titi_phononpattern}). When $E_{\textrm{dep}}$ = $\epsilon$ $\cdot (\cos^{2}(\theta))$ / 2, we see far fewer reflections, and many more instances of energy deposition along a single line. The $E_{\textrm{dep}}$ = $\epsilon \cdot (1 + \cos^{2}(\theta)$) / 2 traces have similar. \\

\section{Discussion}
We have compared the results of two mean free path schemas (the edge absorption and the reflective absorber model) with a geometrical law for athermal phonon propagation. The mean free path schemas only produce consistent results if one assumes a mean free path $\lambda$ larger than the disc radius, and that athermal phonons reflect at the disc borders. The reflective absorber model and the geometrical modelling results reproduce the distribution of the `fast' amplitude $A_{1}$ seen in the experimental data. The reflective absorber model produces a simple exponential decay which then rises again to a second peak at high simulated amplitudes. The three geometrical models also produce similar results, with the energy schemas preferring shallow angles relative to the bismuth, or the scheme preferring angles parallel to the bismuth, producing the most data-like results compared with an energy deposition which is not angle-dependent. \\

Both of these models only work when one accounts for phonon reflection on disc boundaries. Our previous experimental analysis, which had assumed that phonons thermalise at the border of the disc, did not account for this. From this, we can infer that the previous model of assuming that athermal phonons thermalise in the border is incorrect.\\

Our findings are supported by comparing the simulated amplitudes with those from the data, as we have done previously. In both the reflective absorber schema and the geometrical schema we find that the shape of the distribution is similar for a $r_{\textrm{sens}}$ of 600 $\mu$m, which is consistent with the actual thermistor size when one accounts for the simulated shape (disc) compared with the actual shape (rectangular).\\

Finally, a post-facto analysis of related literature of phonon imaging of diamond \cite{hurley1984ballistic} supports the assertion that boundary scattering in the diamond surface is present in the athermal phonon behaviour, and that point sources of energy can generate strong directional variations in the propagation of athermal phonons (supporting the method of Monte Carlo modelling for probing data features). W.M. Gańcza et al. \cite{GANCZA1995423} further demonstrated the viability of Monte Carlo methods for simulating the propagation properties of athermal phonons, alebit with more complex analysis including lattice attributes and quasimomentum. Finally, \cite{jasiukiewicz1992phonon} carried out detailed experimental work in producing the phonon focusing patterns in GaAs heterostructures, and the phonon patterns bear a striking resemblance to the thermalisation pattern shown above, lending credibility to our interpretation.\\

\section{Conclusions}
We have presented three simulation schemas for assessing the mechanisms of athermal phonon propagation in the diamond absorber disc of a composite semiconductor bolometer Bolo 184. In order to probe these properties, we have compared the distribution of fast amplitudes arising from athermal phonon thermalisation from previous experiments \cite{stever2019towards}$^{,}$\cite{stever2019characterisation} with those simulated by our models.\\

The first model assumes phonon propagation and thermalisation determined by a mean free path $\lambda$ smaller than the diameter of the absorber disc, and where phonons arriving at the disc border are absorbed in that location. This model failed to reproduce the distribution shown in the data, in particular by favouring border locations and creating an unrealistic 'build up' at high values of displacement from the disc centre. The second model assumes that the borders of the absorber are reflective and that $\lambda$ is larger than the size of the disc, and produces an amplitude distribution more consistent with the experimental data. Finally, assuming that the phonon thermalisation is determined by a simple geometrical law, rather than a mean free path, also produces consistent results; there are no significant departures in the distributions when one assumes a directionality to the distribution of energy in a 3D space.\\

From this, we conclude that athermal phonons in the diamond absorber of Bolo 184 exhibit reflective properties and a large mean free path, and that this is the dominant effect in the position dependence noted in prior work.\\

\acknowledgments 
The author wishes to acknowledge the French national space agency (Centre national d'études spatiales - CNES) for the doctoral funding allowing for this work to have been carried out, and Dr. Noël Coron for his priceless advice and support.

\bibliography{report} 
\bibliographystyle{spiebib} 

\end{document}